# Optimization of retina-like illumination patterns in ghost imaging

JIE CAO,[1,2,*] DONG ZHOU,[1] YING-QIANG ZHANG,[1] HUAN CUI,[1] FANG-HUA ZHANG,[1] AND QUN HAO,[1,*]

[1] *School of Optics and Photonics, Beijing Institute of Technology, Key Laboratory of Biomimetic Robots and Systems, Ministry of Education, Beijing 100081, China*
[2] *Yangtze Delta Region Academy, Beijing Institute of Technology, Jiaxing 314003, China*
*\* ajieanyyn@163.com, qhao@bit.edu.cn*

**Abstract:** Ghost imaging (GI) reconstructs images using a single-pixel or bucket detector, which has the advantages of scattering robustness, wide spectrum and beyond-visual-field imaging. However, this technique needs large amount of measurements to obtain a sharp image. There have been a lot of methods proposed to overcome this disadvantage. Retina-like patterns, as one of the compressive sensing approaches, enhance the imaging quality of region of interest (ROI) while not increase measurements. The design of the retina-like patterns determines the performance of the ROI in the reconstructed image. Unlike the conventional method to fill in ROI with random patterns, we propose to optimize retina-like patterns by filling in the ROI with the patterns containing the sparsity prior of objects. This proposed method is verified by simulations and experiments compared with conventional GI, retina-like GI and GI using patterns optimized by principal component analysis. The method using optimized retina-like patterns obtain the best imaging quality in ROI than other methods. Meanwhile, the good generalization ability of the optimized retina-like pattern is also verified. While designing the size and position of the ROI of retina-like pattern, the feature information of the target can be obtained to optimize the pattern of ROI. This proposed method paves the way for realizing high-quality GI.

## 1. Introduction

Multi-pixel detectors are usually used as light detection device in conventional optical imaging systems. However, multi-pixel detectors may become expensive or impractical at some specific wavebands, such as the infrared or deep ultraviolet [1]. Pittman et al. first use entangled photon pairs to demonstrate the feasibility of GI in 1995 [2]. GI, also known as correlated imaging or single-pixel imaging [3-8], provides object information by correlating illumination patterns of known light field distribution and a sequence of light intensity collected by a single-pixel or bucket detector. This unique imaging system makes GI have the advantages of scattering robustness, wide spectrum and beyond-visual-field imaging and have been adopted in many related fields including three-dimensional imaging [9-12], terahertz imaging [13-15], multispectral imaging [16-18] and scattering medium imaging [19-21].

In conventional GI, a large number of patterns are required to be illuminated on the target for high quality image while costing much time [22]. For obtaining good performance of GI, how to balance imaging quality and efficiency is still a major challenge to be overcome. Most of studies are focused on two aspects, one is the modulation patterns of illumination [23-25], and the other is reconstruction algorithm [26-28]. Nevertheless, the trade-off between efficiency and quality is still a problem in ROI imaging [29]. Inspired by the foveated vision found in the human eye, modulation patterns with retina-like structure are proposed to obtain higher imaging quality of ROI while not increasing sampling times [29]. In recent years, retina-like patterns have attracted the attention of many scholars. In

2019, Zhang et.al proposed a model of 3D GI combining with retina-like structure to improve imaging efficiency, and some retina-like properties such as invariant scaling and rotation have been realized in proposed imaging system [30]. In 2020, a foveal ghost imaging based on deep learning to realize selecting the ROI for foveal imaging intelligently was proposed by Zhai et al [31]. Selecting the ROI intelligently has been realized in foveal GI by applying generative adversarial networks based on single shot multiBox detector architecture [32]. The high PSNR of the ROI can be achieved compared with uniform-resolution GI. Inspired by human consciousness controlling the eyes to acquire the ROI, Gao et al [33] proposed a new compressive GI called ROI-guided compressive sensing GI which used prior imaging information from fast Fourier single-pixel imaging to achieve a better visual effect and higher imaging quality. Our group presented parallel retina-like computational GI, which is based on a multi-pixel detector combined with retina-like patterns and has better performance than conventional parallel GI and retina-like GI (RGI) [34]. Although the advantages of RGI, to the best of our knowledge, the optimization of illumination retina-like patterns is lack of study. Actually, much attention is paid to the location and size of the ROI or the application of retina-like structure [30-33]. However, there is no study on patterns filled in the ROI of retina-like patterns. The illumination patterns applied to retina-like structure are most random binary patterns. Random patterns are one of the poor performance methods to obtain object information. In this paper, we proposed to improve the imaging quality of RGI by optimizing retina-like patterns, which is realized via filling the patterns containing the sparsity prior of objects in the ROI. The methods to obtain the sparsity prior of objects, such as imaging dictionary and principal component analysis (PCA), have been proposed in previous single-pixel imaging works [35]. We select PCA to generate patterns containing with sparsity prior of objects and then optimize the retina-like patterns to improve the imaging quality of ROI. We demonstrate the optimized retina-like patterns in GI using simulations and experiments. The imaging quality of the proposed scheme is evaluated by comparing the results of conventional RGI.

## 2. Principles

The principle of RGI using optimized patterns by PCA (PCA-RGI) is shown as Fig. 1. A sequence of optimized retina-like patterns is set up on the digital micromirror device (DMD) by computer. After a beam of light from the light source illuminated on the surface of DMD, the reflected light with modulated pattern reaches the object to be imaged. Reflected or projected light illuminated on the object is converged by a focus lens onto the single-pixel detector. The light intensity is collected by a data acquisition and then transmitted to the computer for reconstruction calculation with the corresponding patterns.

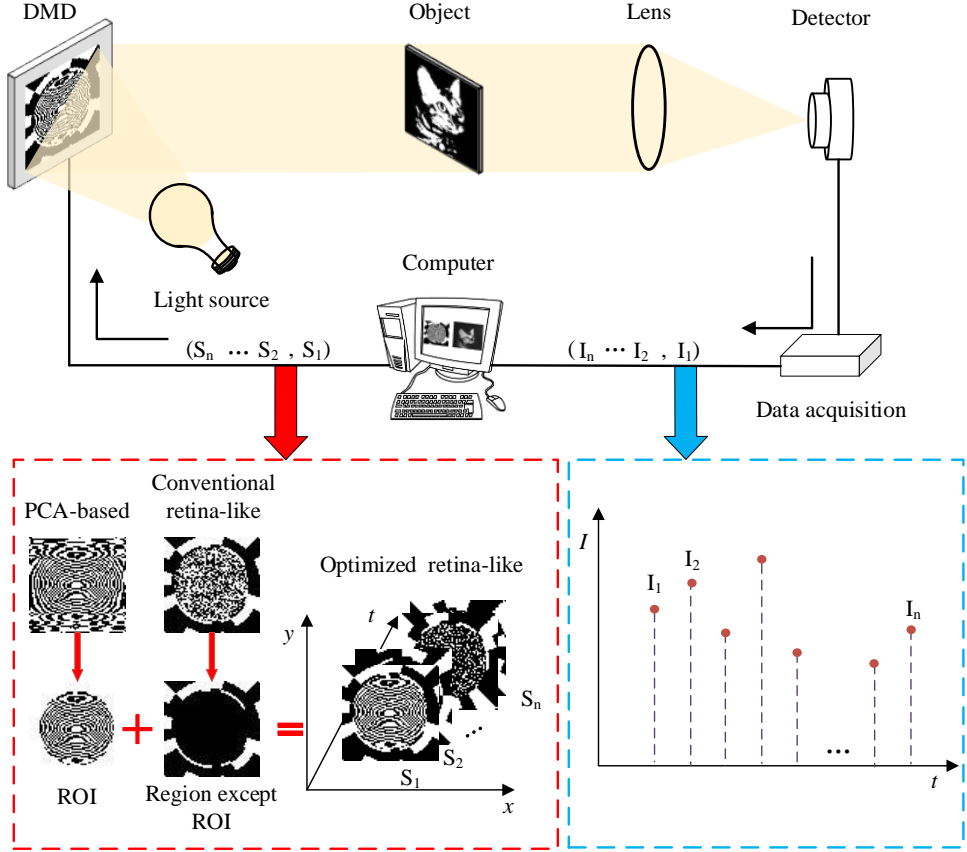

Fig. 1. Principle of PCA-RGI.

The measurements principle of RGI can be described as follows:

$$I_t = \sum_{x,y} S_t(x,y) O(x,y), \tag{1}$$

where $I_t$ represents the light intensity collected by the signal-pixel detector and $t$ is the time index; $O(x, y)$ represents the object and $(x, y)$ represents the 2D Cartesian coordinates in the scene; $S_t(x, y)$ represents the optimized retina-like patterns. The conventional reconstruction algorithm is second-order correlation algorithm. However, compressive sensing (CS) algorithms have been proved to obtain a better performance in reconstructing images than conventional algorithm [36]. In this paper, we select the total variation (TV) regularization prior algorithm [37] for reconstructing images. Unlike the second-order correlation algorithm, the TV-based CS reconstruction algorithm transforms the imaging reconstruction into a constrained optimization problem. The optimization model of TV-based CS is shown as:

$$\begin{aligned}\min \quad & \|c\|_{l_1} \\ s.t. \quad & GO' = c, \\ & S'O' = I'\end{aligned} \tag{2}$$

where $G$ represents the gradient calculation matrix and $c$ is the corresponding coefficient vector; $l_1$ represents the $l_1$ norm; $S' \in R^{T*a}$ represent the optimized retina-like patterns (there are $T$ patterns and each pattern consists of $a = x \times y$ pixels); $O' \in R^{a*1}$ represents the object which is aligned as a vector; $I' \in R^{T*1}$ represents the light intensity.

As for the PCA-based illumination patterns which is used to fill in the ROI of retina-like patterns, we generate them that can be used for DMD in three steps. The first step is to generate grayscale PCA-based patterns from training dataset. Large number of images with features are required to constitute a training dataset. The optimal patterns are then designed based on the common features extracted from the training dataset. It is assumed that dataset contains $M$ objects and $N$ variables. The original training dataset is represented by a matrix X of order $M \times N$, which is shown as:

$$X = \begin{bmatrix} x_{11} & x_{12} & \cdots & x_{1N} \\ x_{21} & x_{22} & \cdots & x_{2N} \\ \vdots & \vdots & \ddots & \vdots \\ x_{M1} & x_{M2} & \cdots & x_{MN} \end{bmatrix}, \quad (3)$$

where each row represents one training image which is converted to a row and $x_{mn}$ represents the value of $n^{th}$ ($n \leqslant N$) pixel in $m^{th}$ ($m \leqslant M$) training image. The raw dataset is composed of many training images and the variables are generally not calculated in the same units. Therefore, the raw dataset needs to be standardized. The standardized process can be expressed as:

$$x_{mn} = (x_{mn} - \langle x_n \rangle) / S_n, \quad (4)$$

where $\langle x_n \rangle$ represents the average value of each column and $S_n$ represents the variance of each column. Then, the covariance matrix $\Sigma$ with a size of $N \times N$ is calculated, which is shown as:

$$\Sigma = \begin{bmatrix} cov(x_1^*, x_1^*) & cov(x_1^*, x_2^*) & \cdots & cov(x_1^*, x_N^*) \\ cov(x_2^*, x_1^*) & cov(x_2^*, x_2^*) & \cdots & cov(x_2^*, x_N^*) \\ \vdots & \vdots & \ddots & \vdots \\ cov(x_N^*, x_1^*) & cov(x_N^*, x_2^*) & \cdots & cov(x_N^*, x_N^*) \end{bmatrix}, \quad (5)$$

where $cov$ represents covariance calculation; $x^*$ represents column vectors in X. The covariance matrix $\Sigma$ is then decomposed into eigenvalues and eigenvectors, which is shown as:

$$Q^T \Sigma Q = \begin{bmatrix} \lambda_1 & & & \\ & \lambda_2 & & \\ & & \ddots & \\ & & & \lambda_N \end{bmatrix}, \quad (6)$$

where $Q$ represents the eigenvector matrix of $\Sigma$ with size $N \times N$ and $\lambda$ represents the eigenvalue of $\Sigma$. Then the eigenvectors in $Q$ are arranged in the descending order of corresponding eigenvalues. Each row vector of $Q$ represents one principal component is also the illumination pattern after conversion to two-dimensional array.

The second step is extracting the pixels with positive value of the PCA-based illumination patterns. The pixel intensity values of the principal components can be both positive and negative. However, in a practical optical system, only positive pixel intensity values can be applied to the projection device. In the previous work [36], both the pixels with positive and negative values are projected. Then the light intensity value corresponding to the original illumination pattern is obtained by subtraction. In our work, we extract the pixels with positive values from the PCA-based illumination patterns and the pixels with negative values are filled with zeros.

The third step is binarizing the grayscale PCA-based illumination patterns. Grayscale illumination patterns cannot be directly loaded into the DMD. There are two typical solutions to binarize the grayscale patterns. One is spatial dithering strategy and the other is temporal dithering strategy [4]. We select the temporal dithering strategy to binarize the grayscale patterns. The grayscale patterns processed by temporal dithering strategy are split into eight binary patterns. Considering the imaging efficiency, we use the lowest bit binary image to replace the original grayscale image.

In the previous work, the ROI of retina-like patterns was filled with random patterns which did not contain any features [30-33]. Unlike conventional methods, we fill in the ROI with PCA-based patterns containing with sparsity prior of objects to optimize the retina-like illumination patterns. We compare the proposed method with the GI using random patterns (Random-GI), GI using conventional retina-like patterns (Random-RGI) and GI using PCA-based patterns (PCA-GI) by simulations and experiments and demonstrate its advantages.

## 3. Simulations and experiments

### 3.1 simulations

Simulations with two different objects were performed separately to evaluate the performance of PCA-RGI compared with Random-GI, Random-RGI and PCA-GI.

The performance of the final reconstructed images was compared quantitatively using the peak signal-to-noise ratio (PSNR) [38] and structural similarity index measure (SSIM) [39] as the evaluation indexes. The PSNR is defined as:

$$\begin{cases} \text{PSNR} = 10\log_{10}\dfrac{(2^k-1)^2}{\text{MSE}} \\ \text{MSE} = \dfrac{1}{a}\sum_{x,y}(O'(x,y)-O(x,y))^2 \end{cases}, \quad (7)$$

where MSE represents the mean square error; $O'(x, y)$ represents the reconstructed image; k is the number of bits set as 8. The higher the PSNR, the better the imaging quality. The SSIM is defined as:

$$\text{SSIM}_{x,y} = \frac{(2\mu\mu'+c_1)(2w+c_2)}{(\mu^2+u'^2+c_1)(\sigma^2+\sigma'^2+c_2)}, \quad (8)$$

where $\mu$ and $\mu'$ represent the average value of $O(x, y)$ and $O'(x, y)$; σ and σ' represent the variance of $O(x, y)$ and $O'(x, y)$; ω represents the covariance between $O(x, y)$ and $O'(x, y)$; $c_1 = (k_1 \times L)^2$ and $c_2 = (k_2 \times L)^2$ are the constant with $k_1 = 0.01$, $k_2 = 0.03$ and $L = 1$. The closer the value of SSIM is to 1, the better the image quality is.

We first clarify several simulation settings. We select 60000 unlabeled grayscale images from STL-10 [40] dataset as the training dataset of PCA-based illumination patterns. The dataset belongs to 10 categories including bird, cat, car and others. We resized all images of dataset to 64×64 pixels which are same as the objects to be imaged. Sampling was simulated by two test images 'coco' and 'lena' under the different measurement conditions. These two images are chosen because coco belongs to the categories of the training dataset, while lena does not. The numbers of the illumination patterns required in the sampling of 10%, 20%, 30%, 40%, 50% and 100% are 410, 819, 1229, 1638, 2048 and 4096. For quantitative comparison, PSNR and SSIM are used to evaluate imaging quality of the overall area and ROI.

The reconstruction results of 'coco' are shown in Fig. 2. It can be seen that the imaging quality of ROI for PCA-RGI is sharper than the other method, especially when the number of measurements is low. Besides, the methods with PCA-based illumination patterns correspondingly obtain better performance than the two methods with random-based patterns.

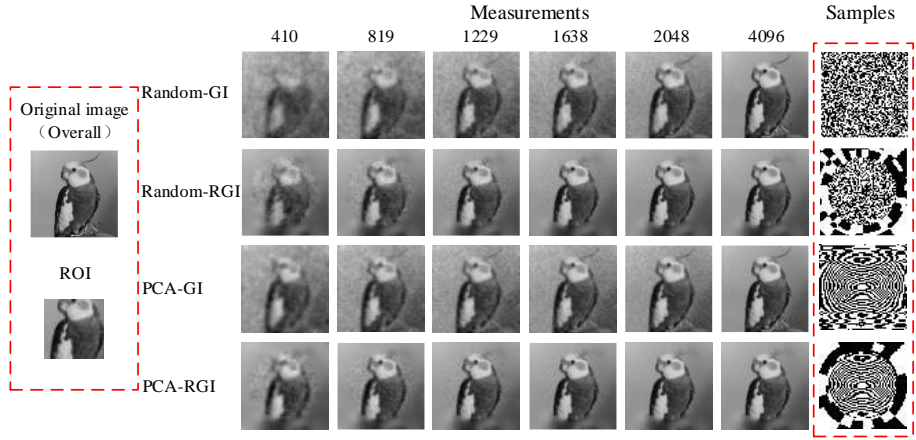

Fig. 2. The comparison results of 'coco' test image reconstucted by Random-GI, Random-RGI, PCA-GI and PCA-RGI under different measurements.

The quantitative comparison results for 'coco' are shown in Table. 1. In terms of the imaging quality of overall image, the PSNR and SSIM of PCA-RGI are better than the other three methods in most conditions. However, we can observe that the PSNR and SSIM of methods with non-retina-like patterns are better than methods with retina-like patterns when the number of measurements is not very low. In terms of the imaging quality of ROI, the PSNR and SSIM of PCA-RGI are always better than the other three methods, followed by PCA-GI, Random-RGI and Random-GI. It is obvious that illumination patterns optimized by PCA have a better performance in obtaining object information than random patterns.

Table 1. Quantitative comparison results of test image 'coco'

| | | Methods | Measurements | | | | | |
|---|---|---|---|---|---|---|---|---|
| | | | 410 | 819 | 1229 | 1638 | 2048 | 4096 |
| PSNR (dB) | Overall | Random-GI | 22.4 | 24.5 | 26.6 | 27.9 | 29.4 | 43.2 |
| | | Random-RGI | 23.4 | 25.9 | 27.4 | 28.2 | 28.7 | 29 |
| | | PCA-GI | 24.7 | 26.5 | **28** | **29.4** | **30.7** | **43.7** |
| | | PCA-RGI | **24.9** | **26.8** | 27.9 | 28.6 | 28.9 | 29 |
| | Fovea | Random-GI | 19.6 | 21.9 | 24.7 | 26.3 | 27.7 | 44.3 |
| | | Random-RGI | 21.1 | 25.4 | 29 | 33 | 37.9 | 50.1 |
| | | PCA-GI | 22.3 | 24.5 | 26.3 | 27.9 | 29.4 | 43.4 |
| | | PCA-RGI | **23.4** | **27.4** | **30.9** | **35** | **41.3** | **64.8** |
| SSIM(A.U.) | Overall | Random-GI | 0.58 | 0.65 | 0.71 | 0.74 | 0.79 | 0.98 |
| | | Random-RGI | 0.7 | 0.79 | 0.84 | 0.87 | 0.89 | 0.91 |
| | | PCA-GI | 0.68 | 0.72 | 0.75 | 0.78 | 0.81 | **0.98** |
| | | PCA-RGI | **0.75** | **0.81** | **0.85** | **0.88** | **0.9** | 0.91 |
| | Fovea | Random-GI | 0.57 | 0.7 | 0.81 | 0.86 | 0.87 | 0.99 |
| | | Random-RGI | 0.65 | 0.81 | 0.9 | 0.95 | 0.98 | 1 |
| | | PCA-GI | 0.7 | 0.8 | 0.85 | 0.88 | 0.9 | 0.99 |
| | | PCA-RGI | **0.74** | **0.86** | **0.92** | **0.96** | **0.99** | **1** |

In order to verify whether the optimized PCA-based retina-like illumination patterns own good generalization ability, the simulation was carried out on the test image 'lena', which not belongs to the training dataset. The settings of the second simulation are same as the first one. The comparison results for 'lena' test image are shown in Fig. 3. It can be observed that the image quality of ROI for PCA-RGI is sharper than the other three methods, which is same as the results of 'coco'. The optimized retina-like illumination patterns also perform better in terms of quickly acquiring information about the object.

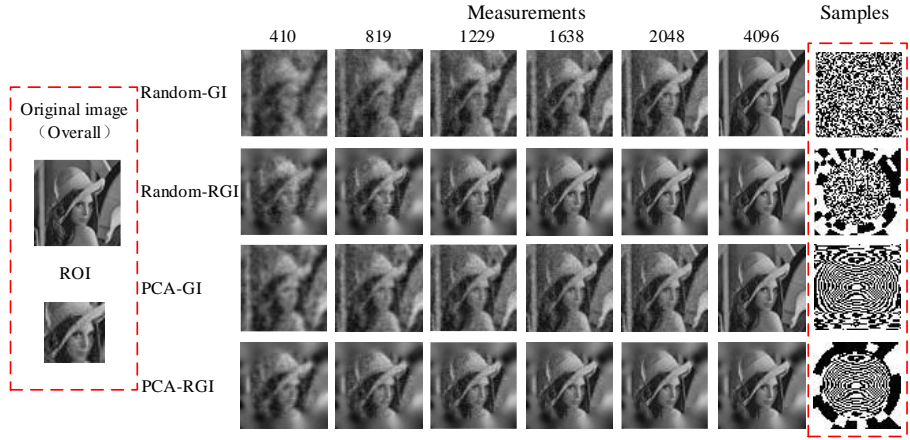

Fig. 3. The comparison results of 'lena' test image reconstructed by Random-GI, Random-RGI, PCA-GI and PCA-RGI under different measurements.

The quantitative comparison results of 'lena' are shown in Table. 2. In terms of the imaging quality of overall image, the PSNR and SSIM of PCA-GI are better than the other three methods which is different with the results of 'coco' for more background information in the region of uninterest. However, in terms of the imaging quality of ROI, the PSNR and SSIM of PCA-RGI are better than the other three methods, followed by PCA-GI, Random-RGI and Random-GI. Obviously, the method with optimized retina-like illumination patterns has better performance because it contains sparsity prior information of the object, even the object does not belong to the category of training dataset.

Table 2. Quantitative comparison results of test image 'lena'

| | | Methods | Measurements | | | | | |
| --- | --- | --- | --- | --- | --- | --- | --- | --- |
| | | | 410 | 819 | 1229 | 1638 | 2048 | 4096 |
| PSNR (dB) | Overall | Random-GI | 19.8 | 21.8 | 23.6 | 24.8 | 26.8 | 40.2 |
| | | Random-RGI | 20.1 | 20.9 | 21.5 | 21.8 | 22 | 22.1 |
| | | PCA-GI | **21.6** | **23.2** | **24.9** | **26.2** | **27.6** | **40.6** |
| | | PCA-RGI | 20.4 | 21.2 | 21.5 | 21.9 | 22 | 22.1 |
| | Fovea | Random-GI | 19.4 | 21.1 | 23 | 24.2 | 26.2 | 40 |
| | | Random-RGI | 21.2 | 24.1 | 27.2 | 30.6 | 35.2 | 48.6 |
| | | PCA-GI | 21 | 22.3 | 23.9 | 25.3 | 26.6 | 40.2 |
| | | PCA-RGI | **22.0** | **25.1** | **27.8** | **32.1** | **39** | **61** |
| SSIM(A.U.) | Overall | Random-GI | 0.53 | 0.65 | 0.74 | 0.79 | 0.85 | 0.99 |
| | | Random-RGI | 0.57 | 0.64 | 0.69 | 0.72 | 0.73 | 0.74 |
| | | PCA-GI | **0.64** | **0.73** | **0.8** | **0.83** | **0.87** | **0.99** |
| | | PCA-RGI | 0.6 | 0.67 | 0.7 | 0.72 | 0.74 | 0.75 |
| | Fovea | Random-GI | 0.54 | 0.64 | 0.75 | 0.81 | 0.88 | 0.99 |
| | | Random-RGI | 0.65 | 0.8 | 0.9 | 0.95 | 0.98 | 1 |
| | | PCA-GI | 0.62 | 0.72 | 0.8 | 0.85 | 0.88 | 0.99 |
| | | PCA-RGI | **0.69** | **0.84** | **0.91** | **0.96** | **0.99** | **1** |

We compare the quantitative results of 'coco' and 'lena' in ROI. We used the PSNR and SSIM of PCA-RGI minus the PSNR and SSIM of random-RGI to express the increment of imaging quality by applying optimized retina-like patterns. The comparison results are shown in Fig. 4, it can be observed that the increment of PSNR and SSIM of 'coco' is more than that of 'lena'. This verifies that the method with optimized PCA-based retina-like patterns obtain better performance for object reconstruction of the same category of the training dataset. Meanwhile, the method also has generalization ability, although its increment is not as good as the objects of training category.

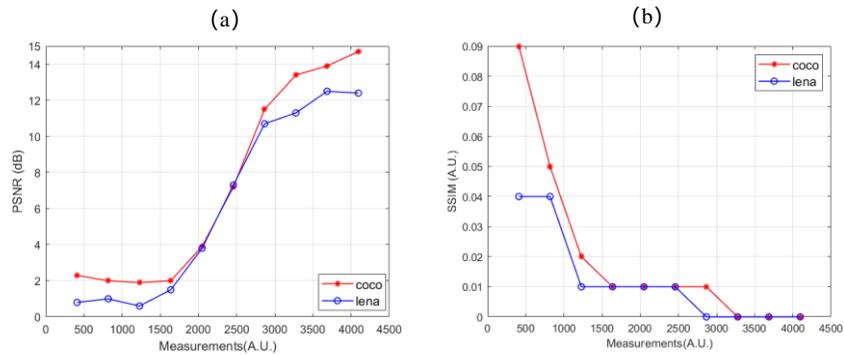

Fig. 4. Quantitative comparison results for quantitative results of 'coco' and 'lena' in ROI. (a) The increment of PSNR of 'coco' and ''lena' is obtained by subtracting the PSNR of Random-RGI from the PSNR of PCA-RGI. (b) The increment of SSIM of 'coco' and ''lena' is obtained by subtracting the SSIM of Random-RGI from the SSIM of PCA-RGI.

In practical optical systems, measurements are always corrupted with noise from ambient light and circuit current. In the above simulations, noise was not considered in the reconstruction. We performed simulations on the influence of measurement noise and compare the robustness of methods with different illumination patterns in ROI. To simulate conditions of different noise levels, we add white Gaussian noise to the measurements of light intensity. The built-in function wgn() of MATLAB was used to add the white Gaussian noise. The reconstructed results of the "coco" with measurements of 2048 and 4096 are considered as original images. The comparison results are shown in Fig. 5, we can observe that the imaging quality of different methods decreases to different degrees with the increase of power of noise. In terms of the results of measurements with 2048, methods with non-retina-like patterns have a better performance than methods with retina-like patterns when power of noise is high. In terms of the results of measurements with 4096, methods with retina-like patterns have a better performance than methods with non-retina-like patterns. Quantitative comparison results are shown in Fig. 6, PCA-RGI has the best performance with the measurements of 2048 and 4096, but not for the condition that measurements are 2048 with the power of noise is more than -10dBw. However, the PCA-RGI always has a better performance than Random-RGI. The robustness of method with retina-like patterns is worse than methods with non-retina-like patterns under the condition measurement is low and power of noise is high enough. The reason is retina-like structure enhances the performance of ROI at the expense of the non-ROI information and also enhances the noise at the same time.

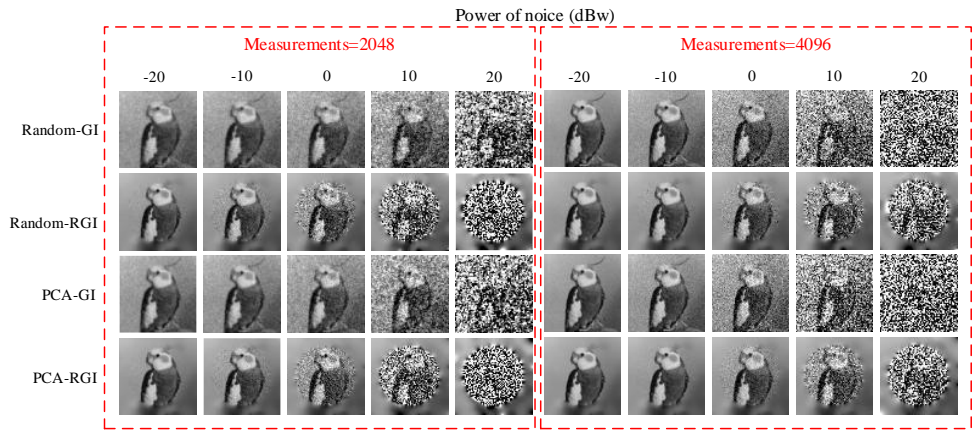

Fig. 5. The noise Comparison results of 'coco' test image in ROI reconstructed by Random-GI, Random-RGI, PCA-GI and PCA-RGI under different power of noise.

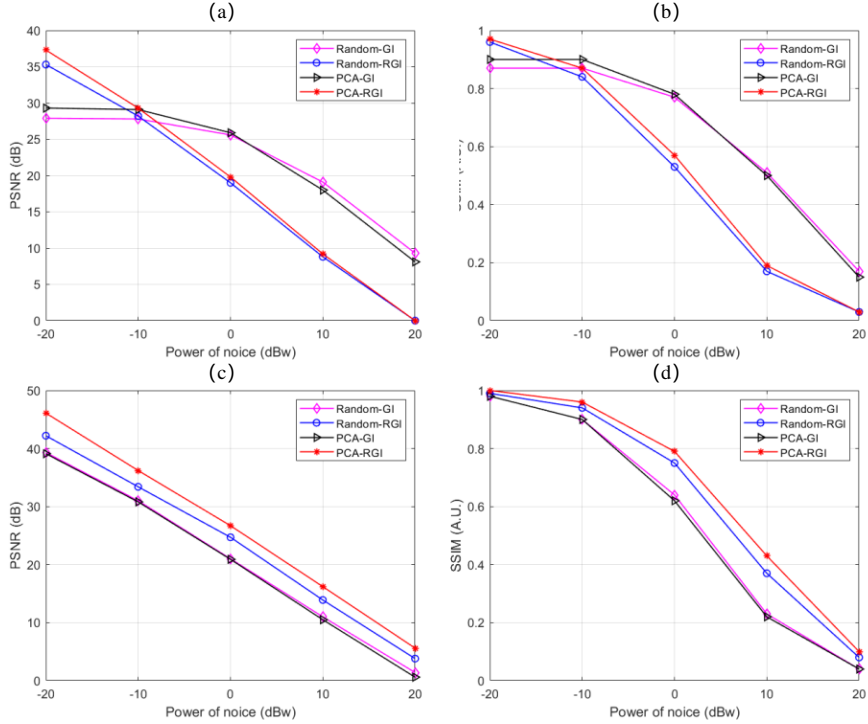

Fig. 6. The quantitative comparison results of 'coco' test image in ROI reconstructed by Random-GI, Random-RGI, PCA-GI and PCA-RGI under different power of noise. (a) The comparison of PSNR with the measurements of 2048. (b) The comparison of SSIM with the measurements of 2048. (c) The comparison of PSNR with the measurements of 4096. (d) The comparison of SSIM with the measurements of 4096.

*3.2 Experiments*

The experimental setup is shown in Fig. 7. The setup consists of illumination part, detection part and objects. The illumination part consists of a light-emitting diode, DMD (Texas Instruments DLP Discovery 4100 development kit) and lens. The light-emitting diode operates at 400-760nm (@20W). The maximum binary modulation rate of DMD is up to 22kHz. The focal length of the projection lens is 150mm. The detection part consists of one photodetector (Thorlabs PDA36A, active area of 13 mm$^2$), data acquisition board (PICO6404E, sampling at 1 MS/s) and a computer. We selected three objects for the experiments, a modified United States Air Force resolution (USAF) test chart, an image of a cat, and an image of 'BIT'. The PCA-based patterns are trained by STL-10 dataset. The image of a cat belongs to the categories of training dataset and the others do not. The experiments were conducted in an environment with ambient noise.

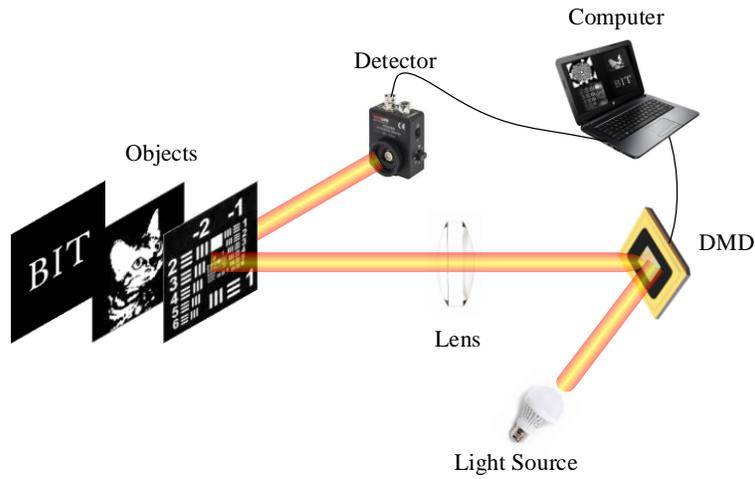

Fig. 7. Experimental setup.

We first used the image of a cat for imaging. Measurements are set as 410, 819, 1229, 1638, 2048 and 4096. The experimental results are shown in Figs. 8 and 9, in terms of imaging of ROI, we can observe that the imaging quality of PCA-RGI and Random-RGI improved as the measurements increased. PCA-RGI has the best performance, followed by PCA-GI, Random-GI and Random-RGI. Affected by the ambient noise, the performance of Random-RGI was greatly affected when the measurements are low. Also affected by noise are Random-GI and PCA-GI, when close to full sampling, the image quality will begin to decrease. PCA-based patterns obviously improve the ability to better obtain information in ROI than other methods. This validates the superiority of optimized retina-like illumination pattern.

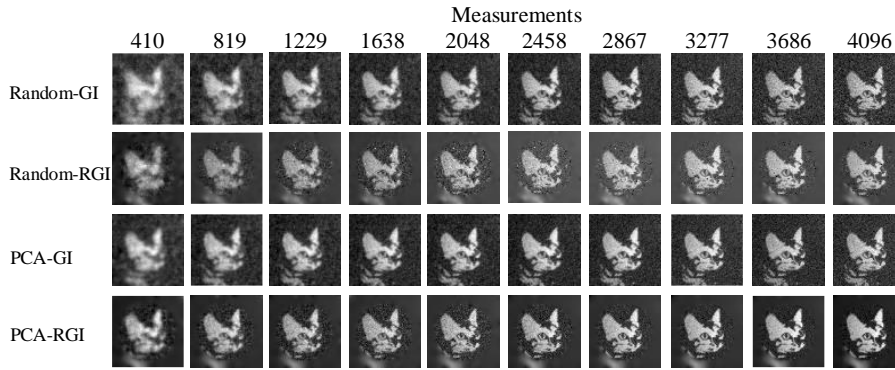

Fig. 8. Experimental results of 'CAT' image reconstruction by Random-GI, Random-RGI, PCA-GI and PCA-RGI under different measurements.

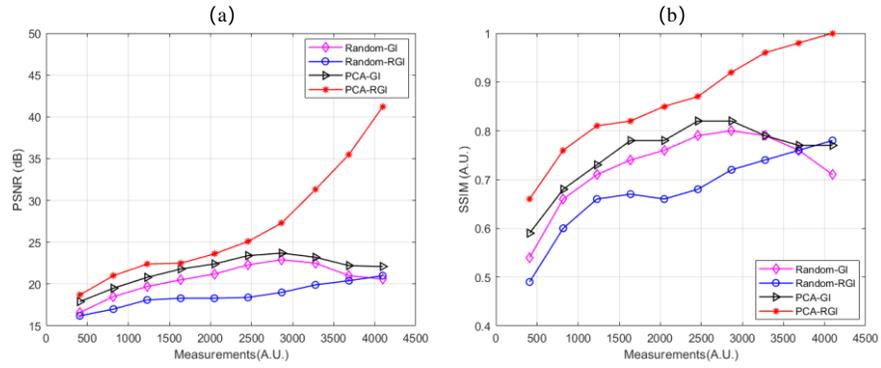

Fig. 9. Quantitative comparison results of 'CAT' image in ROI reconstruction by Random-GI, Random-RGI, PCA-GI and PCA-RGI under different measurements.

The experimental results of 'USAF' and 'BIT' image are shown in Figs. 10 and 11. It can be observed that PCA-RGI has the best imaging quality in ROI. The imaging quality Random-RGI influenced by noise and is even worse than Random-GI. The imaging quality of PCA-RGI is also affected by noise, but its robustness is better than Random-RGI.

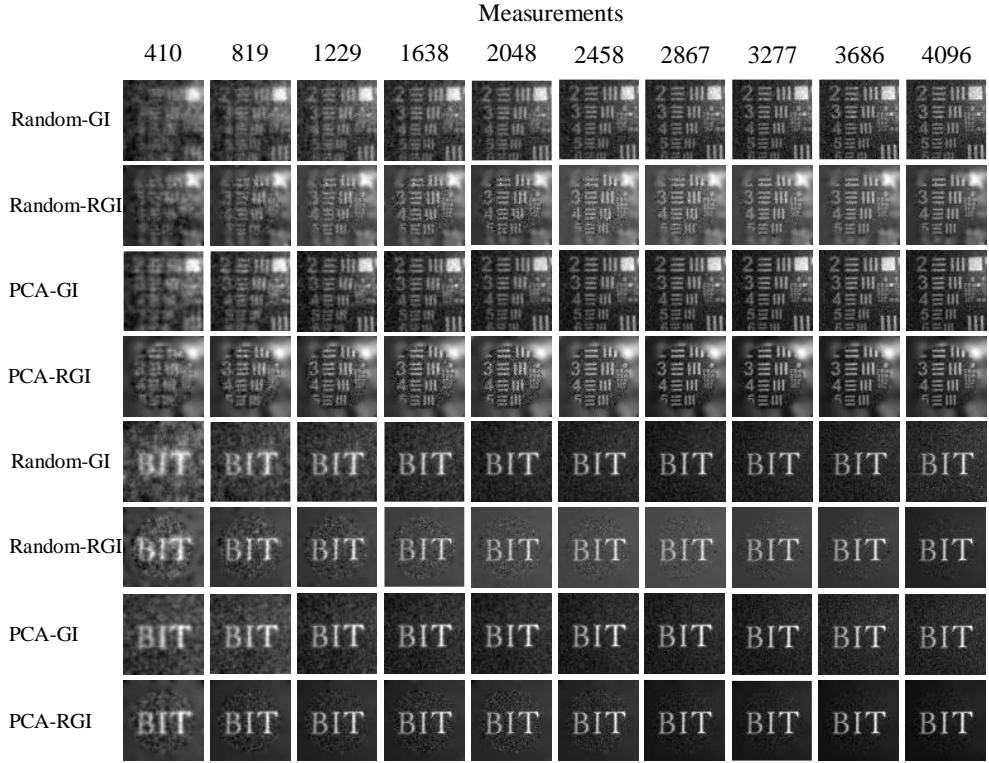

Fig. 10. Experimental results of 'USAF' and 'BIT' image reconstruction by Random-GI, Random-RGI, PCA-GI and PCA-RGI under different measurements.

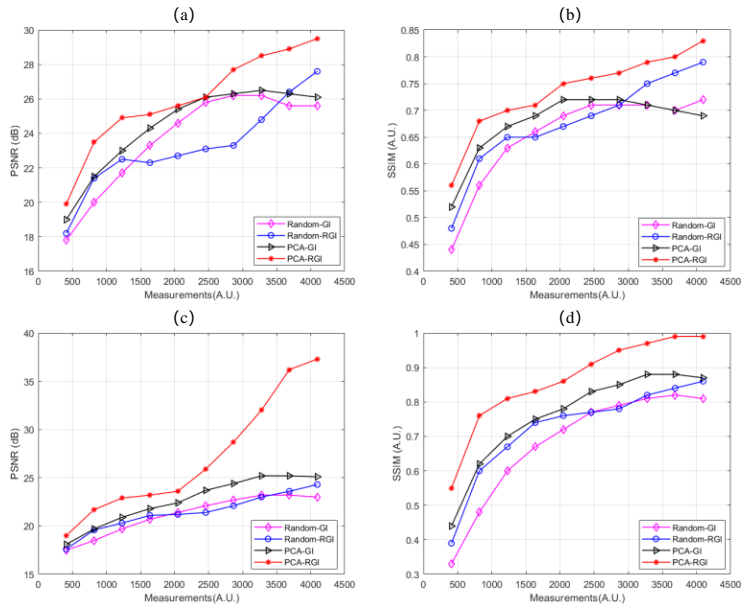

Fig. 11. Quantitative comparison results of 'USAF' and 'BIT' image in ROI reconstruction by Random-GI, Random-RGI, PCA-GI and PCA-RGI under different measurements.

We compared the quantitative results of 'CAT', 'USAF' and 'BIT' in ROI. The quantitative comparison results are shown in Fig. 12, the increment of imaging quality of 'CAT' is better than that of the other two objects. Because PCA-based illumination retina-like patterns improve more for objects belonging to the training dataset. Therefore, although our proposed method with optimized retina-like illumination patterns has better imaging quality in the ROI for the objects of the training dataset category. This method also has good generalization ability.

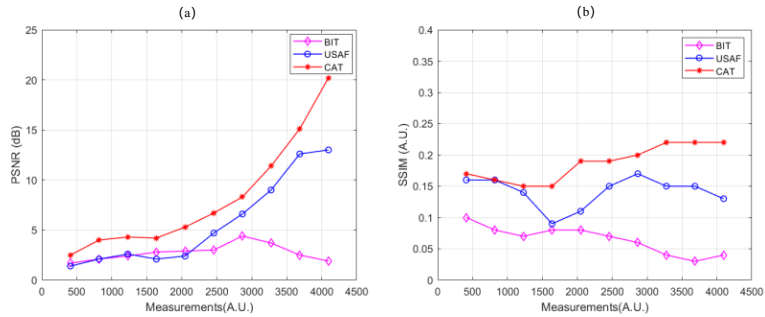

Fig. 12. Quantitative comparison results for quantitative results of 'CAT', 'USAF' and 'BIT' in ROI. (a) The increment of PSNR of 'CAT', 'USAF' and 'BIT' is obtained by subtracting the PSNR of Random-RGI from the PSNR of PCA-RGI. (b) The increment of SSIM of 'CAT', 'USAF' and 'BIT' is obtained by subtracting the SSIM of Random-RGI from the SSIM of PCA-RGI.

## 4. Discussions and Conclusions

The conventional retina-like patterns fill in the ROI with random patterns which has bad performance in obtaining object information. Illumination patterns with sparsity prior of objects can be obtained by performing PCA. In this paper, we present to optimize the retina-like illumination pattern in ROI by PCA. The optimized retina-like patterns are generated by filling the ROI with patterns trained by PCA and keeping the rest of conventional random retina-like patterns. The results of simulations and experiments suggest that PCA-RGI with optimized retina-like illumination patterns has better imaging quality in ROI than Random-GI, Random-RGI and PCA-GI. Meanwhile, PCA-RGI has good generalization ability. For objects that do not belong to the training dataset, PCA-RGI also has the ability to improve the imaging quality of ROI, and the improvement is more significant for objects that belong to the training dataset. However, PCA only can obtain patterns with features of dataset, which is similar to the function of one convolution layer in the deep learning network. Therefore, the patterns of ROI can be more improved by deep learning or other methods to add more priori knowledge. In the current applications of RGI, usually a small amount of sampling is performed to obtain the location and size of the ROI, and then the design and projection of the retina-like patterns is performed. While obtaining the size and location of the target, it can be classified to obtain features, so that adding prior knowledge to the design of the retina-like patterns can significantly improve the imaging quality of ROI. Our proposed method to optimize retina-like patterns paves the way for obtaining high-performance GI.